\documentclass[12pt]{article}
\renewcommand{\thefootnote}{\fnsymbol{footnote}}

\def\ibid#1#2#3{{\it ibid.{} }{\bf #1} (#2) #3}
\def\ie{\hbox{\it i.e.}{}}

\def\nn{\hspace{2mm}}
\def\sss{\scriptscriptstyle}
\newcommand{\MeV}{\mbox{\rm MeV}}
\newcommand{\GeV}{\mbox{\rm GeV}}
\newcommand{\eV}{\mbox{\rm eV}}
\def\sleq{\raisebox{-.6ex}{${\textstyle\stackrel{<}{\sim}}$}}

\def\Tr{{\rm Tr}{}}
\def\Tilde#1{\widetilde{#1}}

\def\sVEV#1{\left\langle #1\right\rangle}

\def\abs#1{\left| #1\right|}

\def\AGUT{{}\;\;\raisebox{.9ex}{$\times$}\raisebox{-.5ex}%
{$\!\!\!\!\!\!\!\!\sss i=1,2,3$} \,(SMG_i \times U(1)_{\sss B-L,i})}%

\begin{document}
\begin{titlepage}
\begin{flushleft}
\vspace*{-1.5cm}
\vbox{\halign{#\hfil        
\cr
DESY 02-066    \cr
NBI-HE-02-07   \cr
hep-ph/0205180 \cr
May 2002  \cr
}}  
\end{flushleft}
\vspace*{-1cm}
\begin{center}
{\Large {\bf Five Adjustable Parameter Fit of Quark and Lepton 
Masses and Mixings}}

\vspace*{7mm}
{\ H. B. Nielsen}\footnote[1]{E-mail: hbech@mail.desy.de}
and {\ Y. Takanishi}\footnote[2]{E-mail: yasutaka@mail.desy.de}
            
\vspace*{.2cm}
{\it Deutsches Elektronen-Synchrotron DESY, \\
Notkestra{\ss}e 85,\\
D-22603 Hamburg, \\
Germany}\\
\vskip .15cm
{\it and}
\vskip .15cm
{\it The Niels Bohr Institute,\\
Blegdamsvej 17, \\
DK-2100 Copenhagen {\O}, \\
Denmark}\\
\vspace*{.3cm}
\end{center}
\begin{abstract}

We develop a model of ours fitting the quark and lepton masses 
and mixing angles by removing from the model a Higgs field previously 
introduced to organise a large atmospheric mixing angle for neutrino 
oscillations. Due to the off-diagonal elements dominating in 
the seesaw neutrino mass matrix the large atmospheric mixing 
angle comes essentially by itself. It turns out that we 
have now only five adjustable Higgs field 
vacuum expectation values needed to fit all the masses and 
mixings {\em order of magnitudewise} taking into account 
the renormalisation group runnings in all sectors. The CHOOZ 
angle comes out close to the experimental bound. 

\end{abstract}
\end{titlepage}
\newpage
\renewcommand{\thefootnote}{\arabic{footnote}}
\setcounter{footnote}{0}
\setcounter{page}{1}

\section{Introduction}
\indent
In previous models~\cite{FN,FNT} based on a series of approximately 
conserved quantum numbers taken to be gauged 
quantum numbers broken by Higgs fields with ``small'' expectation 
values (VEVs) we managed to fit to typically better than a factor two
-- on the average $\pm 60\%$ -- all the masses and mixing angles of quarks 
and leptons known so far. This we did  by means of six adjustable Higgs VEVs 
(not counting the Weinberg-Salam Higgs, which is fixed from 
the Fermi constant) assuming that all coupling constants are of order of 
magnitude unity, only Higgs fields having VEVs of 
orders of magnitude different from unity at Planck scale. The 
adjustable Higgs 
field VEVs were put up to break a series of gauge 
symmetries -- it is enough for the model prediction purpose 
to consider abelian gauge fields, actually two for each 
family of quarks and leptons -- and one of them, $\phi_{\sss B-L}$, 
fits the overall seesaw neutrino scale, while 
two, $\rho$ and $\omega$, connect the first and second family 
by breaking the to these families associated four $U(1)$ gauge groups
down to only two. Now, however, there were in the previous version
of our model three Higgs fields, $W$, $T$ and $\chi$, which in a 
similar way broke two linear combinations of the $U(1)$ gauge groups 
associated with the families number $2$ and $3$. But that is 
seemingly one field too many. It is the main purpose of the 
present article to point out that the model can be improved 
by removing from existence one of those, last mentioned three 
fields, namely the one called $\chi$. For making this avoidance 
of the field, $\chi$, to fit we only need to make an adjustment 
of the quantum numbers of the seesaw scale giving Higgs 
field $\phi_{\sss B-L}$, which we in the new version 
call $\phi_{\rm\sss SS}$.

Our newest modification of removing the field, $\chi$, is really only of 
essential significance for the neutrino oscillation parameters because  
this field only occurred in very insignificant matrix elements for the mass 
matrices of the charged quarks and leptons. By fitting now to the neutrinos 
with fewer parameters in a different way there will of course come a slightly 
different pull for the parameters than before also in the charged sector, 
though.

\section{The model}
\indent
The backbone of the model is the assignment of the in Table~$1$
described quantum numbers under the group 
\begin{equation}
SMG\times U(1)^5\nn,
\end{equation}
supposed to be conceived of as the gauge group assigning two $U(1)$ gauge 
groups to each family of quarks and leptons. In our scheme with three 
seesaw neutrinos, a family consists of a usual Standard Model family 
enriched by one right-handed neutrino (to be used as seesaw 
neutrino). For each family the one of these $U(1)$ gauge groups couples just 
like to the weak hypercharge, $\ie$, with $y/2$ coupling, but {\em only to 
the family in question}. The other one couples to the $B-L$ =
``baryon number''-``lepton number'' again only for the 
{\em family in question}. That is to say we speculate of 
there being altogether six $U(1)$ gauge fields, but now 
one of them -- actually the diagonal subgroup 
of the three of them -- {\em is} the weak hypercharge gauge group
for the whole system and thus we only have added five such groups. 

The model can quite naturally be embellished by the addition to the group
of two extra $SU(2)$ and two extra $SU(3)$ making the whole Standard Model 
group including a $B-L$ charge, which is gauged, become replicated into
one copy for each family of quarks and leptons. So each family
gets its own gauge fields of all the types in the Standard Model
and in addition a gauged $B-L$ charge also separate for each 
family. However, we do not need this embellishment for the mass 
protections that provide the small hierarchies, which we fit. For 
that purpose the abelian quantum numbers in Table~$1$ are 
sufficient. The embellished model has the gauge group, 
\begin{equation}
\AGUT\nn,
\end{equation}
which we call the family replicated gauge group.

The quantum numbers of the fermions and of the Higgs fields which we
have chosen to have in our by now a bit simplified model -- because 
of the deletion of one Higgs field -- are listed in Table~$1$. 
Here the charges, $y_i/2$, are the weak hypercharges for 
the three families $i=1,2,3$, while the Baryon number minus lepton 
number charges $B-L$ for the three families separately are denoted 
$(B-L)_i$, also with $i=1,2,3$.     

\begin{table}[!t]
\caption{All $U(1)$ quantum charges in the family replicated model. 
The symbols for the fermions shall be considered to mean
``proto''-particles. Non-abelian representations are given by a rule 
from the abelian ones (see section $2$).}
\vspace{3mm}
\label{Table1}
\begin{center}
\begin{tabular}{|c||c|c|c|c|c|c|} \hline
& $SMG_1$& $SMG_2$ & $SMG_3$ & $U_{\sss B-L,1}$ & 
$U_{\sss B-L,2}$ & $U_{\sss B-L,3}$ \\ \hline\hline
$u_L,d_L$ &  $\frac{1}{6}$ & $0$ & $0$ & $\frac{1}{3}$ & $0$ & $0$ \\
$u_R$ &  $\frac{2}{3}$ & $0$ & $0$ & $\frac{1}{3}$ & $0$ & $0$ \\
$d_R$ & $-\frac{1}{3}$ & $0$ & $0$ & $\frac{1}{3}$ & $0$ & $0$ \\
$e_L, \nu_{e_{\sss L}}$ & $-\frac{1}{2}$ & $0$ & $0$ & $-1$ & $0$ 
& $0$ \\
$e_R$ & $-1$ & $0$ & $0$ & $-1$ & $0$ & $0$ \\
$\nu_{e_{\sss R}}$ &  $0$ & $0$ & $0$ & $-1$ & $0$ & $0$ \\ \hline
$c_L,s_L$ & $0$ & $\frac{1}{6}$ & $0$ & $0$ & $\frac{1}{3}$ & $0$ \\
$c_R$ &  $0$ & $\frac{2}{3}$ & $0$ & $0$ & $\frac{1}{3}$ & $0$ \\
$s_R$ & $0$ & $-\frac{1}{3}$ & $0$ & $0$ & $\frac{1}{3}$ & $0$\\
$\mu_L, \nu_{\mu_{\sss L}}$ & $0$ & $-\frac{1}{2}$ & $0$ & $0$ & 
$-1$ & $0$\\
$\mu_R$ & $0$ & $-1$ & $0$ & $0$  & $-1$ & $0$ \\
$\nu_{\mu_{\sss R}}$ &  $0$ & $0$ & $0$ & $0$ & $-1$ & $0$ \\ \hline
$t_L,b_L$ & $0$ & $0$ & $\frac{1}{6}$ & $0$ & $0$ & $\frac{1}{3}$ \\
$t_R$ &  $0$ & $0$ & $\frac{2}{3}$ & $0$ & $0$ & $\frac{1}{3}$ \\
$b_R$ & $0$ & $0$ & $-\frac{1}{3}$ & $0$ & $0$ & $\frac{1}{3}$\\
$\tau_L, \nu_{\tau_{\sss L}}$ & $0$ & $0$ & $-\frac{1}{2}$ & $0$ & 
$0$ & $-1$\\
$\tau_R$ & $0$ & $0$ & $-1$ & $0$ & $0$ & $-1$\\
$\nu_{\tau_{\sss R}}$ &  $0$ & $0$ & $0$ & $0$ & $0$ & $-1$ \\ 
\hline \hline
$\phi_{\sss\rm WS}$ & $0$ & $\frac{2}{3}$ & $-\frac{1}{6}$ & $0$ & 
$\frac{1}{3}$ & $-\frac{1}{3}$ \\
$\omega$ & $\frac{1}{6}$ & $-\frac{1}{6}$ & $0$ & $0$ & $0$ & $0$\\
$\rho$ & $0$ & $0$ & $0$ & $-\frac{1}{3}$ & $\frac{1}{3}$ & $0$\\
$W$ & $0$ & $-\frac{1}{2}$ & $\frac{1}{2}$ & $0$ & $-\frac{1}{3}$ 
& $\frac{1}{3}$ \\ 
$T$ & $0$ & $-\frac{1}{6}$ & $\frac{1}{6}$ & $0$ & $0$ & $0$\\
$\phi_{\sss\rm SS}$ & $0$ & $1$ & $-1$ & $0$ & $2$ & $0$ \\ 
\hline
\end{tabular}
\end{center}
\end{table}

The quantum numbers for the embellished model can if one likes be
constructed rather easily by means of the relation between the non-abelian
representation, $\ie$, of $SU(2)$ and $SU(3)$, and the weak hypercharge
$y/2$ in the Standard Model. One namely simply requires to have the same 
relation valid for each value of the family denoting index $i$ separately. 
That is to say, that the representation of $SU(2)_i$ and $SU(3)_i$ for
an arbitrary particle, quark, lepton, or Higgs, are
simply postulated in our embellished version of the model to be 
the same as that quark or lepton Weyl field in the Standard Model 
that has the value $y/2=y_i/2$ for its weak 
hypercharge (actually, though, we 
count it ${\rm mod}~1$ in the case $\phi_{\rm\sss SS}$). 

This embellished model has the beauty of having the biggest 
gauge groups under some restrictions such as: only transforming the known 
particles and the seesaw neutrinos, not unifying Standard Model 
irreducible representations, and having no anomalies~\cite{FNT,agut}.

\subsection{Mass matrices}
\label{sec:mass}
\indent
With the system of quantum numbers in Table~$1$ one 
can easily evaluate, for a given mass matrix 
element, the numbers of Higgs field VEVs of the different types 
needed to perform the transition between the corresponding left- and 
right-handed Weyl fields. The products of 
Higgs fields needed to achieve the quantum number transitions between 
left- and right-handed fermions are 
after removal of the Higgs field in the previous version of our model,
called $\chi$, completely unique. We do no longer have several possibilities 
of achieving the same transition so that a choice of the one giving the 
dominant matrix element would be needed. Rather it is simple linear algebra
with the charges to compute the needed powers for our different fields. 
The main point of the present type of model is that the order 
of magnitudes of the mass matrix elements are 
determined by the number and order of magnitudes of the Higgs field 
vacuum expectation value factors needed. We namely take all the 
Yukawa couplings (and also other couplings) in our model 
and the masses of particles that are not
mass-protected to be of order unity in some fundamental scale (which
we usually imagine to be the Planck scale, and actually take to be that 
in the renormalisation group runnings of the effective couplings 
in our model). Thereby of course the whole calculation we can do 
will only give order of magnitudewise results! Thus one shall 
imagine that every expression for a mass matrix element in addition 
to the Higgs vacuum expectation value factors should be provided 
with an unknown factor of order unity (which 
we in the calculation take to be a random number of order unity),
then later logarithmically averaging the results. We have, however,
not to complicate the expressions too much left out as to be understood
these factors $\lambda_{ij}^{(q)}$, where 
$q \in{}\{\mbox{up, down, charged lepton, Dirac neutrino, 
Majorana neutrino}\}$, 
of order unity. The 
order unity factors physically represent that all the enormously 
many couplings and masses
of not mass protected particles are realistically unknown so 
that the only realistic assumption about them is that they 
are of order of magnitude of
unity (in fundamental units). In our numerical evaluation 
of the predictions of our model we take the ``order unity 
factors'' just in front of each matrix element in our mass 
matrices as dummy integration variables. In fact we integrate
the logarithm of, say, a predicted mass or mixing angle 
(as at first a function of our $9\cdot4+6=40$ order of unity 
parameters) over the real and imaginary parts of these 
parameters with a weight which is normalised so that the 
corresponding integral of one would be unity. The weight is to be 
chosen so that it emphasises the ``order of one factor''
being indeed of order unity. This integral is then 
used as our prediction for the logarithm of the mass or 
mixing in question. This integration over the real and 
imaginary parts of the ``order unity factors'' is in practice
performed by a Monte Carlo method. Had we simply put the 
order one factor equal to unity, we would in some simple cases have 
got roughly right results, however, in many cases
subdeterminants might become exactly zero by such 
procedure and we would have got severely wrong results --
even order of magnitudewise -- compared to our physical 
picture of the fundamental couplings being in reality not 
exactly unity but {\em only} of such order. 

With the order of unity factor removed our mass matrices look as follows: 

\noindent
the up-type quarks:
\begin{eqnarray}
M_{\sss U} \simeq \frac{\sVEV{(\phi_{\sss\rm WS})^\dagger}}{\sqrt{2}}
\hspace{-0.1cm}
\left(\!\begin{array}{ccc}
        (\omega^\dagger)^3 W^\dagger T^2
        & \omega \rho^\dagger W^\dagger T^2
        & \omega \rho^\dagger (W^\dagger)^2 T\\
        (\omega^\dagger)^4 \rho W^\dagger T^2
        &  W^\dagger T^2
        & (W^\dagger)^2 T\\
        (\omega^\dagger)^4 \rho
        & 1
        & W^\dagger T^\dagger
\end{array} \!\right)\label{M_U}
\end{eqnarray}  
\noindent
the down-type quarks:
\begin{eqnarray}
M_{\sss D} \simeq \frac{\sVEV{\phi_{\sss\rm WS}}}
{\sqrt{2}}\hspace{-0.1cm}
\left (\!\begin{array}{ccc}
        \omega^3 W (T^\dagger)^2
      & \omega \rho^\dagger W (T^\dagger)^2
      & \omega \rho^\dagger T^3 \\
        \omega^2 \rho W (T^\dagger)^2
      & W (T^\dagger)^2
      & T^3 \\
        \omega^2 \rho W^2 (T^\dagger)^4
      & W^2 (T^\dagger)^4
      & W T
                        \end{array} \!\right) \label{M_D}
\end{eqnarray}
\noindent %
the charged leptons:
\begin{eqnarray}        
M_{\sss E} \simeq \frac{\sVEV{\phi_{\sss\rm WS}}}
{\sqrt{2}}\hspace{-0.1cm}
\left(\hspace{-0.1 cm}\begin{array}{ccc}
    \omega^3 W (T^\dagger)^2
  & (\omega^\dagger)^3 \rho^3 W (T^\dagger)^2 
  & (\omega^\dagger)^3 \rho^3 W^4 (T^\dagger)^5\\
    \omega^6 (\rho^\dagger)^3  W (T^\dagger)^2 
  &   W (T^\dagger)^2 
  & W^4 (T^\dagger) ^5\\
    \omega^6 (\rho^\dagger)^3  (W^\dagger)^2 T^4 
  & (W^\dagger)^2 T^4
  & WT
\end{array} \hspace{-0.1cm}\right) \label{M_E}
\end{eqnarray}
\noindent
the Dirac neutrinos:
\begin{eqnarray}
M^D_\nu \simeq \frac{\sVEV{(\phi_{\sss\rm WS})^\dagger}}{\sqrt{2}}
\hspace{-0.1cm}
\left(\hspace{-0.1cm}\begin{array}{ccc}
        (\omega^\dagger)^3 W^\dagger T^2
        & (\omega^\dagger)^3 \rho^3 W^\dagger T^2
        & (\omega^\dagger)^3 \rho^3 W^2 (T^\dagger)^7\\ 
        (\rho^\dagger)^3 W^\dagger T^2
        &  W^\dagger T^2
        & W^2 (T ^\dagger)^7\\  
        (\rho^\dagger)^3 (W^\dagger)^4 T^8 
        & (W^\dagger)^4 T^8 
        & W^\dagger T^\dagger
\end{array} \hspace{-0.1 cm}\right)\label{Mdirac}
\end{eqnarray} 
\noindent %
and the Majorana (right-handed) neutrinos:
\begin{eqnarray}    
M_R \simeq \sVEV{\phi_{\sss\rm SS}}\hspace{-0.1cm}
\left (\hspace{-0.1 cm}\begin{array}{ccc}
(\rho^\dagger)^6 T^6 
& (\rho^\dagger)^3 T^6 
& (\rho^\dagger)^3 W^3 (T^\dagger)^3  \\
(\rho^\dagger)^3 T^6
& T^6 & W^3 (T^\dagger)^3 \\
(\rho^\dagger)^3 W^3 (T^\dagger)^3 & W^3 (T^\dagger)^3 & W^6 (T^\dagger)^{12}
\end{array} \hspace{-0.1 cm}\right ) \label{Majorana}
\end{eqnarray}       

The philosophy of the model is that these mass matrices correspond to 
effective Yukawa couplings to be identified with running Yukawa couplings 
at the fundamental/Planck scale for the Higgs field $\phi_{\sss\rm WS}$ 
in the case of the first three mass matrices and for 
$\phi_{\sss\rm SS}$ in the case of the right-handed neutrino 
mass matrix. Therefore these effective Yukawa couplings 
have in principle to be run down by the beta-functions to the scale of 
observation, see section~\ref{RGE}. It is also important that we include the 
``running'' of the irrelevant operator of dimension $5$ giving the neutrino
oscillation masses. 

The right-handed neutrino couplings -- or mass matrix -- is used for
producing an effective mass matrix for the left-handed neutrinos 
which we after the five dimensional running down mentioned take as the ones 
observed in neutrino oscillations~\cite{seesaw,seesawanwenden}:
\begin{equation}
M_{\rm eff} \simeq M_{\nu}^{D} M_R^{-1} (M_{\nu}^D)^T.
\end{equation}

\subsection{$M_{\rm eff}$ crude calculation}
\indent Taking into account the various orders of magnitude of our fitting 
parameters, the expectation values in vacuum of Higgs fields, it is not 
difficult to perform the calculations crudely and very often only one
or two terms dominate a given quantity. Even the slightly more complicated 
case of the effective matrix for the left-handed neutrinos -- the 
neutrino oscillation mass matrix -- we compute crudely:  
\begin{eqnarray}
\label{Meffcrude}
M_{\rm eff} &\sim& \frac{\sVEV{\phi_{\sss\rm WS}}^2}{4 \sVEV{\phi_{\sss SS}} 
W^6 T^6 \rho^6}
\hspace{0.1cm}
\overbrace{\left(\hspace{-0.1cm}\begin{array}{ccc}
\omega^3 W T^2 & \omega^3 \rho^3 W T^2 & \omega^3 \rho^3 W^2 T^7\\ 
\rho^3 W T^2 &  W T^2 & W^2 T^7\\  
\rho^3 W^4 T^8 & W^4 T^8 & W T
\end{array} \hspace{-0.1 cm}\right)}^{M_\nu^D~{\rm part}}\nn\times\nonumber\\
&&\nn\nn\underbrace{\left (\hspace{-0.1 cm}\begin{array}{ccc}
W^6 & W^6 \rho^3 & \sqrt{2} W^3 T^3 \rho^3  \\
W^6 \rho^3 & W^6 \rho^6 & \sqrt{2} W^3 T^3 \rho^6 \\
\sqrt{2} W^3 T^3 \rho^3 & \sqrt{2} W^3 T^3 \rho^6 
& \sqrt{2} T^{6} \rho^6
\end{array} \hspace{-0.1 cm}\right )}_{M^{-1}_R~{\rm part}}\nn
\underbrace{
\left(\hspace{-0.1cm}\begin{array}{ccc}
\omega^3 W T^2 & \rho^3 W T^2 & \rho^3 W^4 T^8 \\
\omega^3 \rho^3 W T^2 &  W T^2 & W^4 T^8 \\
\omega^3 \rho^3 W^2 T^7 & W^2 T^7 &  W T
\end{array} \hspace{-0.1 cm}\right)}_{(M_\nu^D)^T~{\rm part}}\nonumber\\
&\sim& \frac{\sVEV{\phi_{\sss\rm WS}}^2}{4 \sVEV{\phi_{\sss SS}} 
W^6 T^6 \rho^6} \hspace{0.1cm}
\left(\hspace{-0.1cm}\begin{array}{ccc}
\omega^6 W^8 T^4 
& \sqrt{2} \omega^3 \rho^3 W^8 T^4 
& \sqrt{2} \omega^3 \rho^3 W^5 T^6\\
\sqrt{2} \omega^3 \rho^3 W^8 T^4 
& 2 \rho^6 W^8 T^4 
& 2 \rho^6 W^5 T^6 \\
\sqrt{2} \omega^3 \rho^3 W^5 T^6 
& 2 \rho^6 W^5 T^6 
& \sqrt{2} \rho^6 W^2 T^8
\end{array} \hspace{-0.1 cm}\right) \nn.
\end{eqnarray} 
Here we used the following rules of calculation aimed at taking into 
account the random numbers of order unity not written explicitly, but meant 
to be conceived of as standing as factors in front of each term in the 
mass matrices:
\begin{enumerate}
\item When $p$ terms with the same set of ``VEV-factors'', $\ie$, of the types 
$W$, $T$, $\omega$, $\rho$ ($\phi_{\sss\rm WS}$ and $\phi_{\sss\rm SS}$) 
we imagine random phases in the addition (in the complex plane) 
and take the result as $\sqrt{p}$ times the term that 
were $p$ times repeated. 
\item We only take the dominant or couple of dominant terms with respect 
to the VEV-factors. 
\end{enumerate}

\section{Renormalisation group equations}
\label{RGE}
\indent
{}From the Planck scale down to the seesaw scale or rather from 
where our gauge group is broken down to $SMG\times U(1)_{B-L}$ we use
the one-loop renormalisation group running of the Yukawa coupling constant 
matrices, $Y_{\sss U}$, $Y_{\sss D}$, $Y_{\sss E}$, $Y_{\sss \nu}$
and  $Y_{\sss R}$
(being proportional to the mass matrices $M_{\sss U}$, $M_{\sss D}$, 
$M_{\sss E}$, $M_{\sss\nu}^D$ and $M_{\sss R}$, respectively), and the 
gauge couplings~\cite{NT}:
\begin{eqnarray}
\label{eq:recha}
16 \pi^2 {d g_{1}\over d  t} &\!=\!& \frac{41}{10} \, g_1^3 \nn,\\
16 \pi^2 {d g_{2}\over d  t} &\!=\!& - \frac{19}{16} \, g_2^3 \nn, \\
16 \pi^2 {d g_{3}\over d  t} &\!=\!& - 7 \, g_3^3  \nn,\\
16 \pi^2 {d Y_{\sss U}\over d  t} &\!=\!& \frac{3}{2}\, 
\left( Y_{\sss U} (Y_{\sss U})^\dagger
-  Y_{\sss D} (Y_{\sss D})^\dagger\right)\, Y_{\sss U} 
+ \left\{\, Y_{\sss S} - \left(\frac{17}{20} g_1^2 
+ \frac{9}{4} g_2^2 + 8 g_3^2 \right) \right\}\, Y_{\sss U}\nn,\\
16 \pi^2 {d Y_{\sss D}\over d  t} &\!=\!& \frac{3}{2}\, 
\left( Y_{\sss D} (Y_{\sss D})^\dagger
-  Y_{\sss U} (Y_{\sss U})^\dagger\right)\,Y_{\sss D} 
+ \left\{\, Y_{\sss S} - \left(\frac{1}{4} g_1^2 
+ \frac{9}{4} g_2^2 + 8 g_3^2 \right) \right\}\, Y_{\sss D}\nn,\\
16 \pi^2 {d Y_{\sss E}\over d  t} &\!=\!& \frac{3}{2}\, 
\left( Y_{\sss E} (Y_{\sss E})^\dagger
-  Y_{\sss \nu} (Y_{\sss \nu})^\dagger\right)\,Y_{\sss E} 
+ \left\{\, Y_{\sss S} - \left(\frac{9}{4} g_1^2 
+ \frac{9}{4} g_2^2 \right) \right\}\, Y_{\sss E} \nn,\\
\label{Diracyukawa}
16 \pi^2 {d Y_{\sss \nu}\over d  t} &\!=\!& \frac{3}{2}\, 
\left( Y_{\sss \nu} (Y_{\sss \nu})^\dagger
-  Y_{\sss E} (Y_{\sss E})^\dagger\right)\,Y_{\sss \nu} 
+ \left\{\, Y_{\sss S} - \left(\frac{9}{20} g_1^2 
+ \frac{9}{4} g_2^2 \right) \right\}\, Y_{\sss \nu} \nn,\\
16 \pi^2 {d Y_{\sss R}\over d  t} &\!=\!& \left( (Y_{\sss \nu})^\dagger 
Y_{\sss \nu} \right)\,  Y_{\sss R} \,+\, Y_{\sss R} \,\left( (Y_{\sss \nu})^\dagger Y_{\sss \nu} \right)^T\nn,\\
 \label{YScon} Y_{\sss S} &\!=\!& {\Tr}(\, 3\, Y_{\sss U}^\dagger\, Y_{\sss U} 
+  3\, Y_{\sss D}^\dagger \,Y_{\sss D} +  Y_{\sss E}^\dagger\, 
Y_{\sss E} +  Y_{\sss \nu}^\dagger\, Y_{\sss \nu}\,) \nn,
\end{eqnarray}
where $t=\ln\mu$ and $\mu$ is the renormalisation point.

However, below the seesaw scale the right-handed neutrino
are no more relevant and the Dirac neutrino terms in the 
renormalisation group equations should be removed, and also 
the Dirac neutrino Yukawa couplings themselves are not accessible
anymore. That means that, from the seesaw scale down to the 
weak scale, the only leptonic Yukawa beta-functions
should be changed as follows:
\begin{equation}
16 \pi^2 {d Y_{\sss E}\over d  t} =\frac{3}{2}\, 
\left( Y_{\sss E} (Y_{\sss E})^\dagger \right)\,Y_{\sss E} 
+ \left\{\, Y_{\sss S} - \left(\frac{9}{4} g_1^2 
+ \frac{9}{4} g_2^2 \right) \right\}\, Y_{\sss E} \nn.
\end{equation}

Note that the quantity, $Y_{\sss S}$, must be also changed
below the seesaw scale:
\begin{equation}
\label{eq:Y_S}
Y_{\sss S}={\Tr}(\, 3\, Y_{\sss U}^\dagger\, Y_{\sss U} 
+  3\, Y_{\sss D}^\dagger \,Y_{\sss D} +  Y_{\sss E}^\dagger\, 
Y_{\sss E}\,)  \nn.
\end{equation}

Really we stopped the running down according to formula 
(\ref{Diracyukawa}) differently for the different matrix 
elements in the $Y_\nu$ matrix corresponding to the right-handed 
neutrino mass supposed most important for the matrix element 
in question.

Starting the running in an analogous way, we further should evolve 
the effective neutrino mass matrix considered as a five dimensional 
non-renormalisable term~\cite{5run} from the different right-handed 
neutrino masses to the weak scale\footnote{We take the weak scale 
as $180~\GeV$, for simplicity. At this scale we calculated
the pole mass of top quark, too, using $M_t = m_t(M)\left(1
+\frac{4}{3}\frac{\alpha_s(M)}{\pi}\right)$.} 
($180~\GeV$) depending on the terms:
\begin{equation}
\label{eq:remeff}
16 \pi^2 {d M_{\rm eff} \over d  t}
= ( - 3 g_2^2 + 2 \lambda + 2 Y_{\sss S} ) \,M_{\rm eff}
- {3\over 2} \left( M_{\rm eff}\, (Y_{\sss E} Y_{\sss E}^\dagger) 
+ (Y_{\sss E} Y_{\sss E}^\dagger)^T \,M_{\rm eff}\right) \nn,
\end{equation}
where $Y_{\sss S}$ defined in Eq.~(\ref{eq:Y_S}) and in this energy range
the Higgs self-coupling constant running equation is
\begin{equation}
\label{eq:rehiggs}
16 \pi^2 {d \lambda\over d  t}
= 12 \lambda^2 - \left( \frac{9}{5} g_1^2 + 9 g_2^2 \right) \,\lambda
+ \frac{9}{4} \left( \frac{3}{25} g_1^4 
+ \frac{2}{5} g_1^2 g_2^2 + g_2^4 \right) + 4 Y_{\sss S} \lambda 
- 4 H_{\sss S}\nn,
\end{equation}
with
\begin{equation}
 H_{\sss S} = {\Tr} \left\{ 3 \left(Y_{\sss U}^\dagger Y_{\sss U}\right)^2
 + 3 \left(Y_{\sss D}^\dagger Y_{\sss D}\right)^2 +  
\left(Y_{\sss E}^\dagger Y_{\sss E}\right)^2\right\} \nn.
\end{equation}
The mass of the Standard Model Higgs boson is given 
by $M_H^2 = \lambda \sVEV{\phi_{WS}}^2$ and, for definiteness, we 
take $M_H = 115~\GeV$ at weak scale.

{}From $180~\GeV$ down to $1~\GeV$ -- experimental 
scale\footnote{We set the experimentally observable scale as $1~\GeV$,
thus the charged fermion masses and mixing angles are compared to 
``measurements'' at this scale, except the top pole mass.} 
($1~\GeV$) -- we have evaluated the beta-functions with {\rm only} the gauge
coupling constans. In order to run the renormalisation group
equations, we use the following initial values:
\begin{eqnarray}
U(1):\quad & g_1(M_Z) = 0.462 \nn,\quad & g_1(M_{\rm Planck}) = 0.614  \nn,\\
SU(2):\quad & g_2(M_Z) = 0.651 \nn,\quad & g_2(M_{\rm Planck}) = 0.504 \nn,\\
SU(3):\quad & g_3(M_Z) = 1.22  \nn,\quad & g_3(M_{\rm Planck}) = 0.491 \nn.
\end{eqnarray}
Note that we have ignored the influence of the 
$B-L$ gauge coupling constants; however, this effect should not be 
significant, $\ie$, Planck scale to the seesaw scale 
$(\approx 10^{16}~\GeV)$ is only 
$10^3$ order of magnitude difference. Therefore, it should be good
enough for our order magnitude calculations.

\section{Neutrino oscillations}
\indent
The Sudbury Neutrino Observatory (SNO) collaboration
has reported~\cite{SNOnew} recently the measurement of the neutral 
current of the active $^8$B solar neutrino flux and 
related it to measurements of the day and night solar neutrino energy
spectra and rates. Moreover, they presented improved 
determinations of the charged current and 
neutrino-electron scattering rate. It turns out that, 
global analyses~\cite{global} of solar 
neutrino data -- combination of the SNO results with 
previous measurements from other 
experiments~\cite{SNOfirst,SKDN,SK8B,chlorine,sage,gallex,gno} 
-- have confirmed 
that the Large Mixing Angle MSW (LMA-MSW) solution~\cite{MSW} 
gives the best fit to the data and that the LOW solution is 
allowed now only at $2.5\sigma$ and Small Mixing Angle MSW 
(SMA-MSW) solution at $3.7\sigma$,
respectively. Not only that we ``know'' the 
solution of the solar neutrino puzzle but also that
these recent results tell us that a non-zero CHOOZ mixing 
angle~\cite{CHOOZ} is strongly favoured 
at the $3.3\sigma$ C.L. for the LMA-MSW solution~\cite{cc}.

The best fit values of the mass squared difference 
and mixing angle parameters 
in the two flavour LMA-MSW solution somehow
depends on the analysis method, but we take the following point 
as fit values (see in Table~$2$):
\begin{equation}
 \Delta m^2_\odot=5.0\times 10^{-5}~\eV^2 
~~{\rm and}~~\tan^2\theta_{\odot}=0.34\nn. 
\end{equation}

The atmospheric neutrino parameters are the following according to
the Super-Kamio\-kande results~\cite{SK}:
\begin{equation}
 \Delta m^2_{\rm atm}=2.5\times 10^{-3}~\eV^2 
~~{\rm and}~~\tan^2\theta_{\rm atm}=1.0\nn. 
\end{equation}

\section{Results}
\begin{table}[!t]
\caption{Best fit to conventional experimental data.
All masses are running
masses at $1~\GeV$ except the top quark mass which is the pole mass.
Note that we use the square roots of the neutrino data in this 
Table, as the fitted neutrino mass and mixing parameters, 
in our goodness of fit ($\mbox{\rm g.o.f.}$) definition, 
Eq.~(\ref{gof}).}
\begin{displaymath}
\begin{array}{|c|c|c|}
\hline\hline
 & {\rm Fitted} & {\rm Experimental} \\ \hline
m_u & 4.4~\MeV & 4~\MeV \\
m_d & 4.3~\MeV & 9~\MeV \\
m_e & 1.6~\MeV & 0.5~\MeV \\
m_c & 0.64~\GeV & 1.4~\GeV \\
m_s & 295~\MeV & 200~\MeV \\
m_{\mu} & 111~\MeV & 105~\MeV \\
M_t & 202~\GeV & 180~\GeV \\
m_b & 5.7~\GeV & 6.3~\GeV \\
m_{\tau} & 1.46~\GeV & 1.78~\GeV \\
V_{us} & 0.11 & 0.22 \\
V_{cb} & 0.026 & 0.041 \\
V_{ub} & 0.0027 & 0.0035 \\ \hline
\Delta m^2_{\odot} & 9.0 \times 10^{-5}~\eV^2 &  5.0 \times 10^{-5}~\eV^2 \\
\Delta m^2_{\rm atm} & 1.7 \times 10^{-3}~\eV^2 &  2.5 \times 10^{-3}~\eV^2\\
\tan^2\theta_{\odot} &0.26 & 0.34\\
\tan^2\theta_{\rm atm}& 0.65 & 1.0\\
\tan^2\theta_{\rm chooz}  & 2.9 \times 10^{-2} & \sleq~2.6 \times 10^{-2}\\
\hline\hline
\mbox{\rm g.o.f.} &  3.63 & - \\
\hline\hline
\end{array}
\end{displaymath}
\label{convbestfit}
\end{table}

The calculation using random numbers and performed numerically 
was used to fit the masses and mixing angles to the phenomenological 
estimates by minimising what we call ``goodness of fit'',
\begin{equation}
\label{gof}
\mbox{\rm g.o.f.}\equiv\sum_i \left[\ln\left(
\frac{m_{i \,, {\rm pred}}}{m_{i\,, {\rm exp}}}\right) \right]^2 \nn,
\end{equation} 
a kind of $\chi^2$ for the case that we have only order of magnitude accuracy.
The result of the fitting is presented in Table~$2$.
The results presented there were obtained from the following values of the 
set of Higgs VEVs -- where the Higgs field VEVs for the fields
$\rho$, $\omega$, $T$ and $W$ causing the breaking to the diagonal subgroup 
$SMG\times U(1)_{B-L}$ are quoted with the VEV in ``fundamental units'',
while they are for $\phi_{\sss\rm SS}$ (and the not fitted 
$\phi_{\sss\rm WS}$) given in $\GeV$ units:
\begin{eqnarray}
\label{bestvevs}
&&\sVEV{W}=0.157\nn,\nn \sVEV{T}=0.0766\nn,\nn\sVEV{\omega}=0.244\nn, 
\nn\sVEV{\rho}=0.265\nn, \nonumber\\
&&\sVEV{\phi_{\sss\rm SS}}=5.25\times10^{15}~\GeV\nn,
\nn\sVEV{\phi_{\sss\rm WS}}=246~\GeV\nn. 
\end{eqnarray} 

The results of the best fit, with the VEVs in Eq.~(\ref{bestvevs}), 
are shown in Table~$2$ and the fit has  
$\mbox{\rm g.o.f.}=3.63$. To see a typical error, say average
error, compared to the experimental values we should divide
this value with the number of predictions $(17-5=12)$ and then 
take the square root of it: $\sqrt{3.63/12}=0.55$. This means 
that the $12$ degrees of freedom have each of 
them a logarithmic deviation of about $55\%$, $\ie$, we have 
fitted {\rm all quantities} with a typical error 
of a factor $\exp\left(\sqrt{3.63/12}\right)\simeq1.73$ up or 
down. This agrees with theoretically predicted deviations~\cite{FF}. 
However, our worst fitting value is now the electron mass which 
we predict/fit a factor $3$ too heavy.

We should here emphasise this point: even though we reduced the 
number of free parameters by one -- now only five VEVs -- we 
are able to fit all fermion masses and mixing angles 
with an average factor $1.73$ of deviation 
as in previous work.

Experimental results on the values of neutrino mixing angles 
are usually presented in terms of the function $\sin^22\theta$ 
rather than $\tan^2\theta$ (which, contrary to $\sin^22\theta$, 
does not have a maximum at $\theta=\pi/4$ and thus still varies 
in this region). Transforming from $\tan^2\theta$ variables 
to $\sin^22\theta$ 
variables, our predictions for the neutrino mixing angles become:
\begin{eqnarray}
\label{eq:sinmix}
 \sin^22\theta_{\odot} &\!=\!& 0.66\nn,\\
 \sin^22\theta_{\rm atm} &\!=\!& 0.96\nn, \\
\label{eq:sinmixchooz}
 \sin^22\theta_{\rm chooz} &\!=\!& 0.11\nn.
\end{eqnarray}  
We also give here our predicted hierarchical left-handed neutrino masses
$(m_i)$ and the right-handed neutrino masses $(M_i)$ with mass eigenstate
indices $(i=1,2,3)$:
\begin{eqnarray}
m_1 &\!=\!& 1.4\times10^{-3}~~\eV\nn,\nn\nn M_1 = 1.0\times10^{6}~\GeV\nn,\\
m_2 &\!=\!& 9.6\times10^{-3}~~\eV\nn,\nn\nn M_2 = 6.1\times10^{9}~\GeV\nn,\\
m_3 &\!=\!& 4.2\times10^{-2}~~\eV\nn,\nn\nn M_3 = 7.8\times10^{9}~\GeV\nn.
\end{eqnarray}


Note that our fit of the CHOOZ angel (Eq.~\ref{eq:sinmixchooz}) 
is lying on the borderline of the experimental results which analysed
by two flavour method. However, 
our fit satisfy even $2\sigma$ 
C.L. limit ($\tan^2\theta_{\rm chooz}=3.3\times10^{-2}$) 
based on three flavour analysis~\cite{concha}.

{}For the $CP$-violation parameter, the 
Jarlskog triangle area~\cite{cecilia}, $J_{\sss CP}$, we expect 
our prediction to have a larger uncertainty than for the 
quantities, which like the masses 
are essentially mass matrix elements, because it corresponds 
to a ratio or product of six such quantities and 
thus our prediction,
\begin{equation}
J_{\sss CP, {\rm pred}} = 3.6\times10^{-6}
\end{equation}
compared to the ``experimental'' value
\begin{equation}
J_{\sss CP, {\rm exp}} = (2-3.5)\times10^{-5} \nn,
\end{equation}    
should be considered a success of only $1.5\sigma$ deviation.

{}For the for the neutrinoless double beta-decay relevant specially 
weighted neutrino mass average 
\begin{equation}
\label{eq:mmajeff}
\abs{\sVEV{m}} \equiv \abs{\sum_{i=1}^{3} U_{e i}^2 \, m_i} \nn,
\end{equation}
we obtain the prediction from our fit
\begin{equation}
\label{eq:mmajeffres}
\abs{\sVEV{m}}= 3.5\times10^{-3}~\eV\nn.
\end{equation}

Our proton decay predictions may vary a bit with the philosophy, however,
we in any case predict so long lived protons that it is pretty hopeless 
to see any proton decay. At least we get a proton life time of the order:
\begin{equation}
  \label{eq:protonleben}
  \tau_{(p\to\pi^0\,e^+)} \approx 10^{43}~{\rm years}\nn.
\end{equation}
\noindent
The brenching ratio of $\mu\to e+\gamma$ is performed by~\cite{muegamma}:
\begin{equation}
  {\rm Br}(\mu\to e+\gamma)=\frac{3\,\alpha_{em}}{128\,\pi}\,\left( 
\frac{\Delta m^2_{\odot}}{M_W^2} \right)^2 \, \sin^22\theta_{\odot}\nn,
\end{equation}
where $M_W$ is the mass of the $W^{\pm}$ bosons. We insert our 
predictions from Table~$2$ and Eq.~(\ref{eq:sinmix}) in this 
formula, then we get
\begin{equation}
 {\rm Br}(\mu\to e+\gamma)\sim 10^{-56}\nn.
\end{equation}

\section{Simple relations}
\indent
In the approximation of only taking dominant terms and imagining all the 
quantities extrapolated to the Planck scale -- $\ie$ ignoring 
renormalisation group running and sub-dominant terms -- we can 
get the rather simple relations:
\begin{enumerate}
\item Family degeneracy for charged quarks and leptons:
\begin{equation}
m_b \approx m_{\tau}\nn,\nn m_s \approx m_{\mu}\nn, 
\nn m_u \approx m_d \approx m_e \nn, 
\end{equation}
(However, we avoid the $m_c$ and $m_t$ being degenerate with their families).
\item Factorisation of mixing angles for quarks:
\begin{eqnarray}
V_{cb}\,V_{us}\approx V_{ub}\nn,
\end{eqnarray}
\item Neutrino relations:
\begin{equation}
\label{eq:NBeziehung}
\theta_{\odot} \theta_{\rm atm}\approx \theta_{\sss\rm chooz}\nn,
\nn \left(\frac{\Delta m^2_{\odot}}{\Delta m^2_{\rm atm}}
\right)^{\frac{1}{2}}\approx
\frac{1}{2}\theta_{\rm atm}^2 \nn,\nn \left(\frac{m^2_{1}}{\Delta m^2_{\odot}}
\right)^{\frac{1}{2}}\approx \frac{1}{2}\theta_{\odot}^2 \nn, 
\end{equation} 
Here $m_{1}$ is the mass of the lightest left-handed neutrino.
\item Relations for charged quarks and leptons:
\begin{equation}
m_b^3\approx m_t \,m_s \,m_c \nn,\nn m_t \approx 
\hbox{``Weak scale''} \nn,\nn V_{cb} \approx 
\frac{m_s^2}{m_c m_b}\nn.  
\end{equation}
\item Relations between neutrinos and charged quarks and leptons:
\begin{equation}
\label{eq:NCbeziehung}
\theta_{\rm atm}\approx\sqrt{2}\, \frac{W^3}{T^2}\approx 
\sqrt{2}\,\frac{m_c^2}{V_{cb} m_b^2}\nn,\nn \theta_{\odot} 
\approx \frac{\omega^3}{\rho^3} \approx \frac{m_d^2}{m_s^2 V_{us}^3}\nn. 
\end{equation}
\end{enumerate}

Note that according to Eq.~(\ref{eq:NBeziehung}) the neutrino data 
fit by just by two parameters, $\theta_\odot$ and 
$\theta_{\rm atm}$, and one overall scale, $\sVEV{\phi_{\rm\sss SS}}$.
Moreover, according to Eq.~(\ref{eq:NCbeziehung}) the mixing angles
are given in term of the quark quantities.

\section{Theoretical lesson for the mass matrix}
\indent
We should emphasise that we have obtained the 
rather good fit for all the neutrino quantities --
$\theta_\odot$, $\theta_{\rm atm}$, $\theta_{\rm chooz}$,
$\Delta m_{\odot}^2$ and $\Delta m_{\rm atm}^2$ -- and 
the charged masses and mixing angles using only $5$ free 
parameters, namely, five VEVs of Higgs 
fields. Our effective left-handed neutrino mass 
matrix, Eq.~(\ref{Meffcrude}), is of the form
\begin{equation}
  \label{eq:factrization}
\left ( \begin{array}{ccc}
        \phi_{1}^2 &\phi_{1}\phi_{2} & \phi_{1}\phi_{3}\\
 \phi_{1}\phi_{2} &\phi_{2}^2 & \phi_{2}\phi_{3}\\
 \phi_{1}\phi_{3} &\phi_{2}\phi_{3} & \phi_{3}^2
                        \end{array} \right ) \nn, 
\end{equation}
with different order unity factors, which we call factorised mass 
matrix. It is as a consequence
of this type of form we get a relation (\ref{eq:NBeziehung}).
In fact, the ratio of mass 
squared difference, ${\Delta m_{\odot}^2}/{\Delta m_{\rm atm}^2}$, 
is characterised 
by the fourth power of the atmospheric mixing angle at Planck 
scale. Due to 
the large atmospheric neutrino mixing angles, it is clear that 
this fact -- mass squared difference being order unity -- 
is not true experimentally if the effects of renormalisation
equations are not taken into account 
on the Dirac- and Majorana-sectors from Planck scale to 
seesaw scale, and also non-renomalisable
five dimensional evolution equation from the seesaw scale 
to the weak scale. However, with the running included the ``factorised''
mass matrix is a good candidate for fitting the LMA-MSW solution
due to only moderate hierarchy structure of this solution.

\section{Baryogenesis not so successful}
\indent
Our prediction of baryogenesis in the Fukugita-Yanagida scheme~\cite{FY} 
is unfortunately not so successful: The asymmetry from the decay of 
the lightest of the seesaw neutrinos is way too small to 
produce the $B-L$ to give the baryon number asymmetry 
needed for the light element production in big bang at minute 
scale. However, the asymmetry from the decay of the two 
heavy and approximately degenerate seesaw neutrinos is much 
larger and so our hope would be that they could produce the 
asymmetry needed; however, it turns out that the wash-out by the 
processes involving the lightest right-handed neutrino becomes too large 
and we get too little baryon number at the end because 
of the rather high value of the $\Tilde{m}_1$~\cite{Buchmueller}. 

It must though be said that the $\Tilde{m}$'s are products or 
ratios of three mass matrix elements so that the uncertainty 
is expected to be $\exp(\pm\sqrt{3}\cdot64\%)=\exp(\pm 111\%)$, 
meaning a factor $3$ up or down. Our problem with the rather 
high value of $\Tilde{m}$'s is partly due to the renormalisation 
group corrections without which lower $\Tilde{m}$'s by about a 
factor $3$ would result. We also hope that taking into account 
that different flavours will be washed out with different 
rates~\cite{Barbieri} will lower the wash-out rate at the end. 
Somewhat optimistically this could mean that we could use 
an effective $\Tilde{m}_1$ a factor $3$ smaller.

Taking into account also the uncertainty in other quantities we might
hope for that the baryon number predicted only disagrees 
by about three standard deviations.

The many questions of discussion involved in calculating the baryon number 
we should say that we got pretty tight 
to a viable result, however, only based on the idea that it is 
{\em the heavy seesaw neutrinos} that {\em produced the baryon asymmetry}.

\section{Conclusion}
\indent
We have developed our previous model by deleting from the 
list of assumed Higgs fields with vacuum expectation values 
breaking the assumed gauge group $SMG\times U(1)^5$ or $\AGUT$
a Higgs field, $\chi$, which were already suspected to be unwanted by 
having the same quantum numbers as the combination $W^3(T^\dagger)^9$. 
We found that we can indeed fit to order of magnitude accuracy the 
nine charged quark and lepton masses, the two neutrino mass 
square differences, the five measured mixing angles in addition 
to matching the $CP$-violation with {\em only five Higgs 
field vacuum expectation values being adjusted} to 
order of magnitude accuracy. In addition this fit manages to pass 
the test of several experimental bounds: no neutrinoless double 
beta-decay, no proton decay, and no $\mu\rightarrow e+\gamma$ within 
present experimental accuracy, and most interestingly we predict 
a sufficiently small CHOOZ angle, $U_{3e}$; however, there 
is the prediction that we are actually close to the limit 
so that our model might be realistically falsified by decreasing the 
CHOOZ angle limit by an order of magnitude. 

It is remarkable that we fit such a large number, $17$, of genuinely 
measured quantities with only our $5$ parameters order of 
magnitudewise. Our precise gauge group proposals are 
not uniquely called for in as far as we could at least include 
more or less embellishment with non-abelian groups for the 
different families or replace some of the abelian groups by 
non-abelian, but the idea that different families have different 
``chiral'' quantum numbers may be hard to avoid without throwing 
our good agreement out as being accidental.

\section*{Acknowledgements}
We wish to thank T.~Asaka, W.~Buchm{\"u}ller, P.~Di Bari 
and C.~D.~Froggatt for useful discussions. One of us (Y.T.) would like
to thank M.~C.~Gonzalez-Garcia for discussion of 
the CHOOZ angle limit. H.B.N. thanks 
the Alexander von Humboldt-Stiftung for the Forschungspreis. 
Y.T. thanks DESY for financial support.

\end{document}